\documentclass[aps,prb,amsmath,amssymb,reprint,floatfix,showpacs]{revtex4-1}

\usepackage[]{graphicx}	
\usepackage{bm}			
\usepackage{txfonts}	

\newcommand{\sgn}{\ensuremath{\textrm{sgn}}}
\newcommand{\vm}{\ensuremath{\mathbf{m}}}
\newcommand{\vp}{\ensuremath{\mathbf{p}}}

\begin{document}

%
%
\title{Theory of the power spectrum of spin-torque nanocontact vortex oscillators}

\author{Joo-Von Kim}
\email{joo-von.kim@u-psud.fr}
\author{T. Devolder}
\affiliation{Institut d'Electronique Fondamentale, Univ. Paris-Sud, 91405 Orsay, France}
\affiliation{UMR 8622, CNRS, 91405 Orsay, France}

\date{\today}

%
%
\begin{abstract}
Spin-transfer torques in magnetic nanocontacts can lead to self-sustained magnetization oscillations that involve large-amplitude gyrotropic vortex motion. This dynamics consists of a steady state orbit around the nanocontact, which is made possible because the intrinsic magnetic damping is compensated by spin torques. In this article, we present an analytical theory of the power spectrum of these oscillations based on a rigid-vortex model. The appearance of vortex oscillations in nanocontacts is not associated with a Hopf bifurcation: there is no critical current and the only precondition for steady-state oscillations at finite currents is the existence of a vortex in the system, in contrast with conventional spin-torque oscillators that involve large-angle magnetization precession. The oscillation frequency is found to depend linearly on the applied current and inversely proportional to the orbital radius. By solving the associated Langevin problem for the vortex dynamics, the lineshape and linewidth for the power spectrum are also obtained. Under typical experimental conditions, a Lorentzian lineshape with a current-independent linewidth is predicted. Good quantitative agreement between the theory and recent experiments is shown.
\end{abstract}


\maketitle

%
%

\section{Introduction}
A number of novel phenomena in magnetization dynamics are made possible by spin-transfer torques. These torques arise from the transfer of spin angular momentum between a spin-polarized electrical current and local magnetic moments,~\cite{Berger:PRB:1996,Slonczewski:JMMM:1996} and are most apparent in nanoscale structures in which current densities are large. An interesting example of current-driven dynamics are self-sustained oscillations of magnetization, whereby the torques imparted on the local moments act in part to compensate the intrinsic dissipation associated with the magnetization dynamics. As a result of this compensation, the system can attain a self-sustained oscillatory state in which the moments precess freely. These self-sustained oscillations have been observed in a variety of geometries and multilayer compositions,~\cite{Kiselev:Nature:2003,Rippard:PRL:2004,Mistral:APL:2006,Pribiag:NatPhys:2007,Houssameddine:NatMater:2007,Boulle:NatPhys:2007,Pufall:PRB:2007,Deac:NP:2008,Mistral:PRL:2008,Cornelissen:EPL:2009,Dussaux:NC:2010} and have drawn interest for both their fundamental nature and potential applications as nanoscale radio-frequency oscillators.

One class of spin-torque driven oscillations involve magnetic solitons, such as domain walls~\cite{He:APL:2007,Bisig:APL:2009} and vortices.~\cite{Pribiag:NatPhys:2007,Pribiag:PRB:2009,Yamaguchi:APL:2009,Dussaux:NC:2010} Self-sustained oscillations involving vortices were first brought to light in an experiment on spin-valve nanopillars.~\cite{Pribiag:NatPhys:2007} In this study, the physical dimensions and aspect ratio of the active magnetic ``free'' layer of the spin-valve were chosen such that a magnetic vortex ground state is favorable. It was shown that steady-state vortex oscillations could be induced by spin torques with relatively high quality factors, a dynamical behavior that differs fundamentally from the transient or resonant response studied in other works.~\cite{Novosad:PRB:2002b, Novosad:PRB:2005, Buchanan:NatPhys:2005} Since then, other experiments have shown that similar behavior can be observed in magnetic tunnel junctions.~\cite{Dussaux:NC:2010}

A more striking example of self-sustained vortex oscillations has been observed in magnetic nanocontacts.~\cite{Pufall:PRB:2007,Mistral:PRL:2008} In this geometry, large current densities are attained by channeling electron flow through a magnetoresistive multilayer stack via a metallic point contact, which is typically tens of nanometers in radius.~\cite{Rippard:PRL:2004} In contrast to nanopillars where the physical geometry confines the vortex motion, the magnetic free layer in a nanocontact structure is a continuous film with much larger lateral dimensions (typically tens of microns), so translational invariance for the vortex is restored within the film plane in the absence of structural or magnetic defects. A confining potential in this medium arises, however, from the Zeeman energy due to the Oersted-Amp{\`e}re fields associated with the applied current through the nanocontact. If we imagine the nanocontact to be an infinitely long cylindrical conductor, then the Oersted-Amp{\`e}re fields generated by the current flow follow a circular pattern in the film plane, which shares the same cylindrical symmetry as the magnetic vortex. Because of this, the Zeeman energy due to these fields acts as a confining potential for the vortex around the nanocontact, with the potential minimum being centered on the nanocontact (if this is perfectly circular).
%
\begin{figure}
\includegraphics[width=7cm]{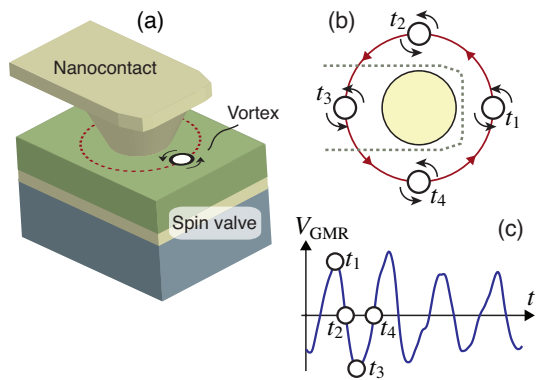}
\caption{\label{fig:schema}(Color online) (a) Schematic illustration of current-induced vortex oscillations in a magnetic nanocontact system. The vortex, assumed to be in the free layer (top) of a spin valve stack, executes a large-radius orbit around the nanocontact in a potential generated by the Oersted-Amp{\`e}re field associated with electrical current flow through the contact. (b) Vortex orbital motion giving rise to a (c) time-varying component of the giant magnetoresistance (GMR) voltage signal.}
\end{figure}
The resulting dynamics in this geometry, as revealed by micromagnetics simulations,~\cite{Mistral:PRL:2008} is a steady state motion of the vortex that involves a large orbit around the nanocontact, as shown schematically in Fig.~\ref{fig:schema}. By virtue of such large orbital motion, the free layer magnetization underneath the contact undergoes full rotations, which in turn results in a large time-varying component of the magnetoresistance. The observation of sub-gigahertz GMR variations in experimental nanocontact systems has been attributed to such vortex dynamics.~\cite{Mistral:PRL:2008, Pufall:PRB:2007, Keller:APL:2009, Devolder:APL:2009, Manfrini:APL:2009, vanKampen:JPD:2009, Devolder:SPIE:2009, Ruotolo:NN:2009}

Nanocontact vortex oscillations represent a dynamical state associated with the current flow, because spatial confinement for the vortex is absent when no current is applied (aside from defects that can pin the vortex). Furthermore, the existence itself of a vortex is not guaranteed in the absence of currents, because the vortex state bears a higher cost in magnetic energy than the uniform ground state. For these reasons, vortex oscillations in nanocontacts differ fundamentally from their counterparts in nanopillars, where the physical geometry guarantees both the vortex state and a confining potential that allow for damped or resonant dynamics. This distinction means that existing theories of spin-torque nano-oscillators, which describe a current-driven Hopf bifurcation between a damped oscillatory state and a limit cycle,~\cite{Rezende:PRL:2005,Slavin:ITM:2005,Kim:PRB:2006,Tiberkevich:APL:2007,Kim:PRL:2008,Tiberkevich:PRB:2008} or other theoretical works describing current-driven vortex dynamics in magnetic dots,~\cite{Ivanov:PRL:2007,Liu:APL:2007,Choi:APL:2008,Moon:PRB:2009,Gaididei:IJQC:2009,Khvalkovskiy:PRB:2009} cannot be applied directly to the nanocontact system.

In this article, we present an analytical theory of the power spectrum of current-driven vortex oscillations in magnetic nanocontacts. While some elements of this theory have been reported in a prior publication,~\cite{Mistral:PRL:2008} a number of experimental observations since have required an extension of that earlier work. In particular, the revised theory presented here goes toward explaining two key experimental facts. First, it is now well established that vortex oscillations in in-plane magnetized systems can be nucleated and sustained without any applied magnetic fields.~\cite{Devolder:APL:2009,Manfrini:APL:2009,vanKampen:JPD:2009} This observation cannot be explained by the earlier theory in which a spin-polarization component perpendicular to the film plane, which is expected to be negligibly small in the absence of applied perpendicular fields, is necessary for sustaining oscillations.~\cite{Mistral:PRL:2008} While this component is known to be important for nanopillars,~\cite{Dussaux:NC:2010} the reproducible nature of the zero-field result for the nanocontact system suggests a different spin-transfer mechanism is predominant for sustaining oscillations. Second, the quality factor of the nanocontact vortex oscillations (i.e., the ratio between the oscillation frequency and the spectral linewidth) is typically an order of magnitude lower than that for nanopillar excitations. This is not immediately intuitive, since high quality factors have been observed for vortex oscillations in nanopillars.~\cite{Pribiag:PRB:2009} Finally, we derive an analytical form for the Zeeman energy associated with the Oersted-Amp{\`e}re field generated by the applied currents and show that it varies linearly with the radial distance for large amplitude oscillations. This linear dependence was used in earlier work, but it was only found empirically from numerical calculations.

This paper is organized as follows. In Section II, we present the geometry, model and equations of motion in the limit of the rigid vortex approximation used. There, we derive the Thiele equation describing the vortex dynamics. In Section III, we consider the large amplitude limit of steady-state vortex motion in which the orbit is far away outside the nanocontact. In this limit, we derive some simplified equations of motion and analytical solutions for the oscillation radius and frequency. In Section IV, we examine the stochastic dynamics in the large orbit limit and derive analytical forms for the spectral linewidth. Some discussion of the results obtained and comparison with experimental data and other work are presented in Section V. A summary of key results is given in Section VI.

\section{Model}

\begin{figure}
\includegraphics[width=8.5cm]{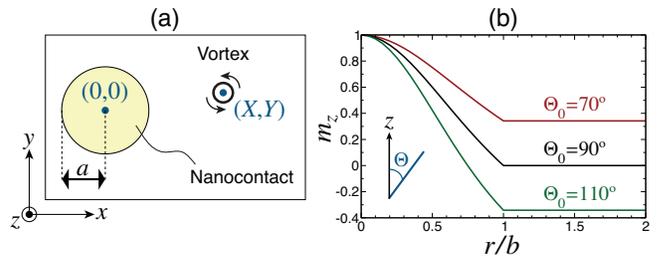}
\caption{\label{fig:geometry_profile}(Color online) (a) Geometry of the nanocontact in the film plane of the free magnetic layer and system of coordinates used. The nanocontact makes a circular cross section with the free layer with radius $a$ and centered at the origin. $(X,Y)$ denotes the position of the vortex core in the film plane. (b) Profile of vortex core for different magnetization tilt angles $\Theta_0$.}
\end{figure}

The theory we present describes the magnetization dynamics of the free magnetic layer in a spin-valve stack, as shown in Fig.~\ref{fig:geometry_profile}a. We do not account for any coupling that may appear between the free and reference magnetic layers, and the dynamics of the latter are ignored. The geometry we consider is defined in Fig.~\ref{fig:geometry_profile}a. In our notation, $z$ represents the axis perpendicular to the film plane, $a$ the nanocontact radius, and $d$ the free layer thickness.
%

\begin{figure}
\includegraphics[width=7cm]{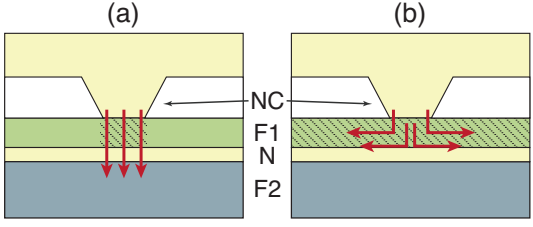}
\caption{\label{fig:cip_cpp}(Color online) Sources of spin-torques in the free ferromagnetic layer (F1), with the arrows indicating current flow. (a) Current perpendicular-to-plane (CPP) and (b) current in-plane (CIP) torques both contribute to vortex dynamics. The hashed regions indicate the regions in which the spin-torques act. NC denotes the nanocontact, N the metallic spacer, and F2 the ferromagnetic reference layer.}
\end{figure}

The magnetization dynamics is governed by the Landau-Lifshitz equation of motion, with an additional phenomenological Gilbert damping term and a spin-torque term $\Gamma_{\rm st}$,
\begin{equation}
\frac{\partial \vm}{\partial t} +|\gamma_0| \vm \times \mathbf{H}_{\rm eff} = \alpha\, \vm \times \frac{\partial \vm}{\partial t} + \mathbf{\Gamma}_{\rm st},
\label{eq:LLG}
\end{equation}
where $\vm$ is a unit vector representing the magnetization orientation, $\gamma_0 = \mu_0 |\gamma|$ is the gyromagnetic constant, $\mu_0 \mathbf{H}_{\rm eff} = -\nabla_{\mathbf{M}}\mathcal{E}$ is the effective field, and $\alpha$ is the Gilbert damping constant. There are two main sources of spin torques that are relevant in this geometry, $\mathbf{\Gamma}_{\rm st} = \mathbf{\Gamma}_{\rm CPP} + \mathbf{\Gamma}_{\rm CIP}$ (see Fig.~\ref{fig:cip_cpp}). 
%
The first involves torques related to electron transport perpendicular to the film plane (CPP), which is associated with the CPP magnetoresistance of the spin valve and affects only the magnetization dynamics \emph{within} the nanocontact region,
\begin{equation}
\mathbf{\Gamma}_{\rm CPP} =  -\sigma_{1} I \; \mathcal{P}(\vm \cdot \vp) \;  \vm \times (\vp \times \vm).
\label{eq:STTCPP}
\end{equation}
The unit vector $\vp$ represents the magnetization orientation of the reference layer and the scalar function $\mathcal{P}$ accounts for the angular dependence of spin transfer, which depends strongly on the layer thicknesses and material parameters of the multilayer stack.~\cite{Slonczewski:JMMM:1996,Xiao:PRB:2004} $I$ is the applied current representing \emph{electron} flow and
\begin{equation}
\sigma_1 =  P_1 \frac{\hbar}{e} \frac{\gamma}{M_s d} \frac{1}{\pi a^2}
\end{equation}
is the CPP spin-transfer efficiency, where $P_1$ is the spin polarization of the CPP current, and $M_s$ is the saturation magnetization of the free layer. Under this convention, $I > 0$ represents the flow of spins (electrons) from the free to the reference layer along the negative $z$ direction. The second spin torque contribution originates from electron transport within the film plane (CIP). Such in-plane currents arise in experimental systems because the electrical pads that connect to the bottom electrode are usually located at a significant lateral distance from the nanocontact. Because electrical currents flowing in a ferromagnetic metal are also spin-polarized, spin-torques arise in regions in which a spatial gradient in magnetization occurs. These CIP torques can be separated into an adiabatic and a nonadiabatic component, respectively,~\cite{Zhang:PRL:2004,Thiaville:EPL:2005}
\begin{equation}
\mathbf{\Gamma}_{\rm CIP} = -(\mathbf{u} \cdot \mathbf{\nabla})\vm + \beta \; \vm \times [(\mathbf{u} \cdot \mathbf{\nabla})\vm],
\end{equation}
which arise from different spin transport regimes relative to the spatial magnetization gradients. In this description, the spin torques are more conveniently parametrized by an effective spin drift current,
\begin{equation}
\mathbf{u} =  P_2 \frac{\hbar}{2e} \frac{\gamma}{M_s} \mathbf{j}(\mathbf{r}), 
\end{equation}
where $\mathbf{j}$ is the current density of electron flow in the film plane and $P_2$ represents the spin polarization of this CIP current.~\footnote{Note that $P_1$ is largely determined by the spin-dependent transport properties across the entire multilayer stack, while $P_2$ depends largely on the transport properties of the magnetic free layer.} In order to keep the ensuing calculations tractable a simple model for electron flow is used, which follows from the schematic illustration presented in Fig.~\ref{fig:cip_cpp}b. The applied current is taken to flow uniformly into the cylindrical region underneath the nanocontact in the free layer, and the subsequent electron flow within the film plane has a uniform density normal to the surface of this cylinder. A similar argument is applied to the flow within the nanocontact. Within this approximation, the current density in the film plane is purely radial from the nanocontact,
\begin{equation}
\mathbf{j}(\mathbf{r}) = \hat{\mathbf{r}} \,\frac{I}{2\pi a d} \times %
\begin{cases}
r /a  & r < a \\
a/ r & r \geq a
\end{cases}
\end{equation}

In what follows, we consider the dynamics of a single magnetic vortex in the free layer. We will not seek to describe the nucleation process here as it is beyond the scope of this work. (A recent experiment has shown that vortices and vortex oscillations can be initiated in a reproducible way in such nanocontact systems~\cite{Devolder:APL:2010}). Instead, we will focus on how the vortex responds under the influence of the spin torques described above. A crucial assumption we make is that the spatial profile of the vortex remains constant throughout its motion. While the vortex core can deform significantly during the course of its motion,~\cite{[{See, e.g., }]Vansteenkiste:NP:2009} and that such changes can be account for by spin-wave theories or through the inclusion of high-order time derivatives in the equations of motion,~\cite{Mertens:2000} such effects will not be treated here for the sake of simplicity.

Let $(\Theta,\Phi)$ represent the magnetization orientation in polar coordinates. In the rigid vortex approximation, the spatial profile can be parametrized entirely in terms of the vortex core position in the film plane, $\mathbf{X} = (X,Y)$, for which the origin is taken to be the center of the nanocontact. We consider a vortex with an angular variation of magnetization in the film plane of the form

\begin{equation}
\Phi(x,y;X,Y) = n \tan^{-1}\left(\frac{y-Y}{x-X}\right) + \frac{\pi}{2},
\end{equation}
where $n=\pm1$ is the topological charge ($n=1$ for a vortex, $n=-1$ for an antivortex). The component of magnetization perpendicular to the film plane, which describes the vortex core profile, is given by a function of the form $\Theta = \Theta(\|\mathbf{r} - \mathbf{R} \|)$, where $\mathbf{r}$ and $\mathbf{R}$ denote the position radial vectors in polar coordinates, i.e., $\|\mathbf{r} - \mathbf{R} \|^2 = (x-X)^2 + (y-Y)^2$. The exact functional form we choose for the vortex core results only in small quantitative  differences in the resulting dynamics. For the sake of consistency for the remainder of this paper, we use a modified version of the Usov ansatz for the vortex core profile that accounts for the uniform tilt angle, $\Theta_0$, of the free layer magnetization out of the film plane,~\cite{Gaididei:IJQC:2009}
\begin{equation}
\cos\Theta(r;R) = %
\begin{cases}
\dfrac{b^2 + (2 \cos\Theta_0 - 1)\|\mathbf{r} - \mathbf{R} \|^2}{b^2 + \|\mathbf{r} - \mathbf{R} \|^2} p, & \|\mathbf{r} - \mathbf{R} \| < b \\
0, & \|\mathbf{r} - \mathbf{R} \| \geq b
\end{cases}
\end{equation}
where $b$ is the vortex core radius and $p= \pm 1$ is the core polarization, which describes the orientation of the core magnetization relative to the normal of the film plane. An illustration of the core profile for three tilt angles $\Theta_0$ is given in Fig.~\ref{fig:geometry_profile}b. The tilt angle is important to describe the magnetization profile under fields applied perpendicularly to the film plane. A detailed comparison between simulated core profiles, this and other ansatz, are given in Ref.~\onlinecite{Gaididei:IJQC:2009}.

The dynamics of the rigid-vortex is derived following the method of collective coordinates.~\cite{Rajaraman:1982} We elevate the core position to a dynamic variable, $\mathbf{X} \rightarrow \mathbf{X}(t)$, which allows us to express the spatial magnetization profile as $\Theta = \Theta[\mathbf{x} - \mathbf{X}(t)]$, with an analogous expression for $\Phi$. The ensuing dynamics is then generated using the spin Lagrangian
\begin{equation}
L = \int dV \; \mathcal{L}(\Theta[\mathbf{x} - \mathbf{X}(t)],\Phi[\mathbf{x} - \mathbf{X}(t)]).
\label{eq:lagrangian}
\end{equation}
By integrating over the Lagrangian density, $\mathcal{L}$, with the chosen magnetization profile for the vortex, we generate the equation of motion for each (generalized) coordinate $\xi$ in the usual way,
\begin{equation}
\frac{d}{dt} \frac{\partial L}{\partial \dot{\xi}} - \frac{\partial L}{\partial \xi} = F_{\xi,\rm nc},
\label{eq:el}
\end{equation}
where $F_{\xi,\rm nc}$ represents the nonconservative forces such as damping and spin-torques.

The conservative part of the Lagrangian dynamics, i.e., the left-hand side of (\ref{eq:el}), is given by the spin Berry phase term, $\mathcal{L}_B = (M_s/\gamma) \dot{\Phi} (1-\cos\Theta)$, and the magnetic energy density that plays the role of a potential, $\mathcal{U} = \mathcal{U}(\Theta,\Phi)$. The Euler-Lagrange equations become
\begin{equation}
\mathbf{G} \times \frac{d \mathbf{X}}{d t} + \frac{\partial U}{\partial \mathbf{X}} = \mathbf{F}_{\rm nc}.
\end{equation}
The first term on the left hand side is the gyrotropic term, where the gyrovector $\mathbf{G}$ is defined by
\begin{equation}
\mathbf{G} = \frac{M_s}{\gamma} \int dV \sin{\Theta} (\nabla\Phi \times \nabla\Theta)
\end{equation}
and is equal to $\mathbf{G} = G \hat{\mathbf{z}}$, where $G = 2 \pi (p - \cos\Theta_0) n d M_s/\gamma$. The second term is obtained directly from the magnetic energy density $U = \int dV \; \mathcal{U}$. If we assume that the lateral dimensions of the film are much larger than the vortex core, the magnetostatic energy, in addition to the usual exchange and crystalline anisotropy energies, becomes independent of the vortex core position. Translational symmetry in the film plane is therefore only broken by the Oersted-Amp{\`e}re field generated by the electrical current applied through the contact, constituting the sole contribution to the potential $U$. For the sake of simplicity, we assume that this Oersted-Amp{\`e}re field is equivalent to that produced by an infinitely long current-carrying cylindrical conductor, which allows the Zeeman energy associated with this field to be written as
\begin{equation}
U_Z = -\mu_0 M_s H_I \int dV f(r) \sin\Theta (\cos\phi \cos\Phi + \sin\phi \sin\Phi),
\label{eq:zeeman}
\end{equation}
where $H_I = |I|/(2\pi a)$ is the magnitude of the Oersted-Amp{\`e}re field at the edge of the contact, $(r, \phi)$ denotes the spatial variables $(x,y)$ in polar coordinates.  The function $f(r)$ describes the variation of the magnitude of the Oersted field as a function of radial distance $r$ in the film plane,
\begin{equation}
f(r) = %
\begin{cases}
r/a & r < a \\
a/r & r \geq a
\end{cases}
\end{equation}

The nonconservative forces we consider are Gilbert damping and spin torques, as discussed above. In a Lagrangian formalism, the force due to Gilbert damping $F_{\rm G}$ can be included  by means of a Rayleigh dissipation function $W$,
\begin{equation}
F_{\xi,\rm G} = -\frac{\partial W}{\partial \dot{\xi}} = - \frac{\alpha M_s}{2 \gamma} \frac{\partial}{\partial \dot{\xi}} \int dV \; \left[ \dot{\Theta}^2 + \sin^2\Theta  \; \dot{\Phi}^2 \right],
\end{equation}
which leads to
\begin{equation}
\mathbf{F}_{\rm G} = -\alpha \overleftrightarrow{\mathbf{D}} \cdot \frac{d \mathbf{X}}{d t},
\end{equation}
where $\alpha$ is the Gilbert damping constant and $\overleftrightarrow{\mathbf{D}}$ is the damping dyadic,
\begin{equation}
\overleftrightarrow{\mathbf{D}} = \frac{M_s}{\gamma} \int dV \left( \nabla\Theta \otimes  \nabla\Theta + \sin^2{\Theta} \nabla\Phi \otimes  \nabla\Phi \right).
\end{equation}
With the core profile chosen, this term is computed to be $\overleftrightarrow{\mathbf{D}} = D \overleftrightarrow{\mathbf{I}}$, where $\overleftrightarrow{\mathbf{I}}$ is the $2 \times 2$ identity matrix, $D = \pi M_s d \left(2 + \sin^2\Theta_0 \ln(L/b) \right)/\gamma$, and $L$ is lateral size of the free layer.

To describe the CPP spin torques, we first identify the relevant forces using the Landau-Lifshitz equation (\ref{eq:LLG}) and (\ref{eq:STTCPP}),
\begin{eqnarray}
F_{\Theta,\rm CPP} & = & 0,\\
F_{\Phi,\rm CPP} & = & -\sigma_{\rm CPP} I \, \frac{M_s}{\gamma}\; \mathcal{P}(\vm \cdot \mathbf{p}) \sin^2 \Theta. 
\end{eqnarray}
By applying the chain rule, we find that the CPP torques can be divided into two components, $\mathbf{F}_{\rm CPP} = \mathbf{F}_{\textrm{CPP},||} + \mathbf{F}_{\textrm{CPP},\perp}$, which describe contributions from the in-plane and perpendicular-to-plane components, respectively, of the spin polarization unit vector $\mathbf{p}$,
\begin{eqnarray}
\mathbf{F}_{\textrm{CPP},||} & = & \sigma_{\rm CPP} I \, \mathcal{P}\frac{M_s}{\gamma} \int_{\rm NC} dV \; p_{||} \, \biggl( \nabla \Theta \, \sin\Phi + \\ \notag 
&& \frac{1}{2}\nabla\Phi \cos{2\Theta} \sin{\Phi}  \biggr), \label{eq:cpp_par} \\
\mathbf{F}_{\textrm{CPP},\perp} & = & -\sigma_{\rm CPP} I \, \mathcal{P} \frac{M_s}{\gamma} \int_{\rm NC} dV \; p_{\perp} \sin^2\Theta \, \nabla\Phi, \label{eq:cpp_perp}
\end{eqnarray}
where the volume integration is limited to the region of the free layer underneath the nanocontact (NC). Without loss of generality, we have assumed that the in-plane spin polarization component is along the $x$ axis. To account for the CIP torques, we note that the adiabatic and nonadiabatic terms can be derived by generalizing the time derivatives to convectional derivatives using the spin-drift velocity $\mathbf{u}$, in both the Landau-Lifshitz and Lagrangian formulations. The adiabatic component can be obtained from the generalized Berry phase term,~\cite{Shibata:PRL:2005}
\begin{equation}
\frac{\partial \Phi}{\partial t}(1-\cos \Theta) \rightarrow \left( \frac{\partial}{\partial t} + \mathbf{u} \cdot \mathbf{\nabla} \right) \Phi \; (1-\cos \Theta),
\end{equation}
which leads to a term that resembles the gyrotropic force,
\begin{equation}
\mathbf{F}_{\rm ad} = -\frac{M_s}{\gamma} \int dV \; \sin\Theta (\nabla\Phi \times \nabla \Theta) \times \mathbf{u},
\label{eq:adforce}
\end{equation}
while the nonadiabatic term requires a proportionality factor $\beta/\alpha$, i.e., we make the substitution 
\begin{equation}
\frac{\partial}{\partial t} \rightarrow \frac{\partial}{\partial t} +  \frac{\beta}{\alpha}\mathbf{u} \cdot \mathbf{\nabla}
\end{equation}
for the time derivatives in the Rayleigh dissipation function, which reflects the fact that Galilean invariance is not generally present for magnetic dissipative processes in real systems.~\cite{Barnes:PRL:2005,Lemaho:PRB:2009} As expected, the resulting form for the nonadiabatic term is similar to Gilbert damping,
\begin{equation}
\mathbf{F}_{\rm n-ad} = -\frac{\beta M_s}{\gamma} \int dV \; \left( \nabla\Theta \otimes  \nabla\Theta + \sin^2{\Theta} \nabla\Phi \otimes  \nabla\Phi \right) \cdot \mathbf{u}.
\label{eq:nadforce}
\end{equation}
%

\section{\label{sec:large_orbit}Steady-state oscillations}

This section is devoted to the description of the large-orbit vortex motion in the steady state. It begins with a brief discussion on the initial transient dynamics leading to the steady state, where it is shown that the parallel component of the CPP torques play a major role in driving the vortex out of the nanocontact. Next, a simple form for the Zeeman energy due to the Oersted-Amp{\`e}re field is found. This result, combined with the CIP torques in the large-amplitude limit, leads to equations of motion that are solved to give the steady state orbital radius and frequency.

For the initial transient dynamics, the small-amplitude limit is considered in which the vortex is assumed to remain within the nanocontact area and close to its center. In this limit, the CPP torques can be readily evaluated. For the purposes of illustrating the qualitative behavior in this limit, it is assumed that $\mathcal{P} = 1$ and no external magnetic field is applied so that the tilt magnetization angle can be ignored, i.e. $\Theta_0 = \pi/2$. The parallel component $F_{\rm{CPP},||}$ is nonvanishing only within the vortex core where the gradient in the polar angle, $\nabla\Theta$, and $\cos(\Theta)$ terms are nonvanishing. As such, it suffices to limit the integration in (\ref{eq:cpp_par}) over only the vortex core region, which leads to
\begin{equation}
\mathbf{F}_{\rm{CPP},||} = \sigma_{1} I \frac{M_s}{\gamma} \, (\pi b d p) \left(\ln{2} \, \hat{\mathbf{i}} + \frac{\pi-2}{4} \, \hat{\mathbf{j}} \right).
\end{equation}
This force is independent of the vortex position and acts to drive the vortex out of the nanocontact area. A good estimate of the perpendicular component of the CPP torques can be obtained by neglecting the core contribution using the approximation $\sin^2\Theta \approx 1$. Within this approximation, the integral in (\ref{eq:cpp_perp}) reduces to a simple integral of the quantity $\nabla \Phi$, which yields
\begin{equation}
\mathbf{F}_{\rm{CPP},\perp} = \sigma_{1} I \frac{M_s}{\gamma}\left(\frac{\pi a^2}{2}\right) \frac{1}{X^2+Y^2} \left( -Y \,\hat{\mathbf{i}} + X \,\hat{\mathbf{j}}  \right) + O(b^2).
\end{equation}
In contrast to the parallel component, the perpendicular CPP spin torques acts to drive a gyrotropic motion for the vortex, with a magnitude that is inversely proportional to the radial vortex distance from the nanocontact center. The combination of the two CPP spin torques gives a force acting on the vortex that leads to a spiraling motion of the vortex out of the nanocontact area.

For the remainder of this section, we will assume that the steady-state regime is attained after such transient dynamics. We focus on the large amplitude motion of the vortex around the nanocontact, which is relevant for describing the power spectrum of the steady-state oscillations as observed experimentally. We will not concern ourselves with the transient dynamics associated with the nucleation process; this is beyond the scope of this paper and will be treated elsewhere.

\begin{figure}
\includegraphics[width=6cm]{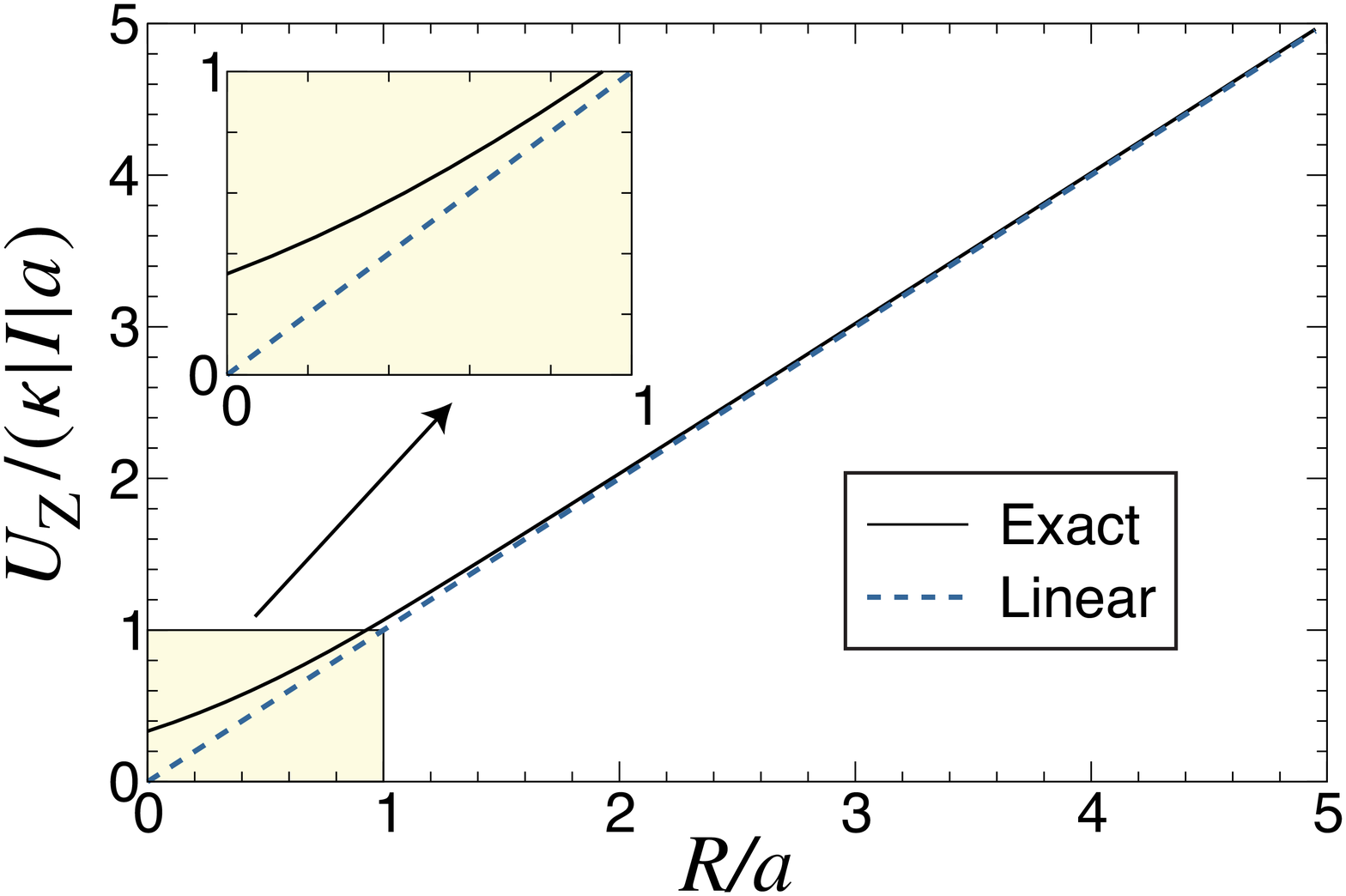}
\caption{\label{fig:zeeman_energy} (Color online) Comparison between the exact form and linear approximation of Zeeman energy, $U_{\rm Z}$, as a function of scaled radial distance $R/a$, where $a$ is the point contact radius. The inset represents a zoom for $R/a \leq 1$.}
\end{figure}

The Zeeman energy due to the Oersted-Amp{\`e}re fields can be computed without difficulty if contributions from the vortex core are neglected; we take $\Theta = \Theta_0$ in (\ref{eq:zeeman}) and retain only the magnetization variation in $\Phi$.  Let $(r, \phi)$ and $(R,\varphi)$ represent the spatial variables $(x,y)$ and $(X,Y)$ in polar coordinates, respectively. We proceed by separating the integration in Eq.~\ref{eq:zeeman} into two parts. The first part involves integrating over the contact area, which is found to be
\begin{equation}
\begin{split}
U_{Z,\rm in} & = \sin(\Theta_0)\int_{0}^{2\pi} d\phi \int_{0}^{a} dr \; \frac{r^2}{a} \frac{r-R\cos{\phi}}{\sqrt{r^2 - 2 r R \cos{\phi}+R^2}}, \\
& =  \sin(\Theta_0)\frac{4 |a-R |}{9a} \bigl[ (2a^2-R^2) E(\psi) + (a+R)^2 K(\psi) \bigr],
\end{split}
\end{equation}
where $\psi \equiv -4 a R / (a-R)^2 $, and $K(\psi) = \int_{0}^{\pi/2} dz \; (1-\psi \sin^2 z)^{-1/2} $ and $E(\psi) = \int_{0}^{\pi/2} dz \; (1-\psi \sin^2 z)^{1/2}$ are elliptic integrals. The second integral, which runs over the region outside the contact, is infinite. Nevertheless, we can identify the infinite background term from the indefinite integral
\begin{multline}
\int_{0}^{2\pi} d\phi \int dr \;  \frac{a( r-R\cos{\phi})}{\sqrt{r^2 - 2 r R \cos{\phi}+R^2}} \\
= 4 a |r-R| E\left[-\frac{4 r R}{(r-R)^2}\right],
\end{multline}
by noting that $E[-4 rR/(r-R)^2] = \pi/2$ as $r \rightarrow \infty$, from which it is deduced that
\begin{equation}
U_{Z,\rm out} = \mu_0 M_s H_I d \sin(\Theta_0) \left[ 2\pi a R + 4 a |a-R| E(\psi) \right],
\end{equation}
with $\psi$ as defined above. In the large amplitude limit the elliptic integrals can be expanded in a power series in terms of $a/R$, which allows the total Zeeman energy $U_Z = U_{Z,\rm in} + U_{Z,\rm out}$ to be expressed as
\begin{equation}
\begin{split}
U_Z &\approx \mu_0 M_s H_I d \sin{\Theta_0} (4 \pi a R), \\
 & = \left(2 \mu_0 M_s d \sin{\Theta_0} \right) |I| R \equiv \kappa(\Theta_0) \, |I| \, R.
\end{split}
\end{equation}
This linear dependence on the radial distance of the vortex was found empirically through numerical calculations in previous work,~\cite{Mistral:PRL:2008} but we provide a solid basis for this functional form here. The linear approximation is very good for $R/a \geq 1$, which can be seen in Fig.~\ref{fig:zeeman_energy} where a comparison with the exact solution is presented.
%

For in-plane magnetized systems, such as the CoFe or Permalloy films studied in experiment,~\cite{Pufall:PRB:2007,Mistral:PRL:2008,vanKampen:JPD:2009,Devolder:APL:2009,Manfrini:APL:2009} we assert that the CPP spin torques can be neglected in the \emph{absence of any applied magnetic fields}. Under these conditions, both in-plane and perpendicular components of the CPP torques should vanish for the following reasons. First, in the absence of any perpendicular fields, the $p_\perp$ component should be vanishingly small because the magnetization of the reference layer, which is either a hard ferromagnetic material or is exchange biased by an antiferromagnet, lies entirely in the film plane. This perpendicular component is important, however, for dynamics in large perpendicular applied fields.~\cite{Mistral:PRL:2008,Dussaux:NC:2010} Second, if the vortex core is sufficiently far from the nanocontact, the magnetization gradient $\nabla \Theta$ vanishes and $\cos 2\Theta \simeq \cos \pi =0$ within the nanocontact region. As a result the parallel component also vanishes, which can be seen by inspecting Eq.~\ref{eq:cpp_par}.

We are therefore led to the conclusion that CIP spin torques are the dominant mechanism for self-sustained vortex oscillations in the large orbit limit under zero or low applied fields. Let $\sigma_2$ represent the spin-torque efficiency for the CIP component in this limit,
\begin{equation}
\sigma_2 \equiv P_2 \frac{\hbar}{e} \frac{\gamma}{M_s d} \frac{1}{4\pi a^2},
\end{equation}
such that the spin-current drift velocity can be expressed as ($r \geq a$),
\begin{equation}
\mathbf{u}(r) = \sigma_2 I \; \frac{a^2}{r}  \hat{\mathbf{r}}.
\end{equation}
With this definition, $\sigma_2 I$ retains units of angular frequency, in accordance with the convention adopted in spin torque oscillation theory. By substituting this functional dependence for the adiabatic torques into (\ref{eq:adforce}) above and integrating over the region outside the nanocontact, we find the simple and appealing result
\begin{equation}
\mathbf{F}_{\rm ad} = \mathbf{G} \times \mathbf{u}(\mathbf{X}).
\end{equation}
Similarly, the nonadiabatic torques (\ref{eq:nadforce}) under the same assumptions are found to be
\begin{equation}
\mathbf{F}_{\rm n-ad} = \beta D \, \mathbf{u}(\mathbf{X}).
\end{equation}

\begin{figure}
\includegraphics[width=6cm]{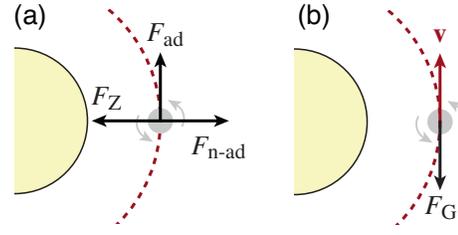}
\caption{\label{fig:forces}(Color online) Schematic diagram illustrating the different force terms in Eq.~\ref{eq:thiele}. (a) Static forces: Zeeman potential ($F_{\rm G}$), adiabatic ($F_{\rm ad}$), and non-adiabatic ($F_{\rm n-ad}$) torques, for $I>0$. (b) Gilbert damping, $F_{\rm G}$, associated with the vortex motion $\mathbf{v}$.}
\end{figure}

The full equation of motion for the current-driven vortex in the large amplitude limit is therefore given by
\begin{equation}
\mathbf{G} \times \left[  \frac{d \mathbf{X}}{d t} - \mathbf{u}(\mathbf{X}) \right] + D \left[ \alpha \frac{d \mathbf{X}}{d t} - \beta \, \mathbf{u}(\mathbf{X})  \right] + \frac{\partial U}{\partial \mathbf{X}} = 0.
\label{eq:thiele}
\end{equation}
The role of each term is shown schematically in Fig.~\ref{fig:forces}.
%
The conic structure of the Zeeman potential leads to a force that pulls the vortex towards the nanocontact center, while the nonadiabatic torque counteracts this force pushing the vortex radially outwards. The adiabatic torque acts as a ``boost'' along the circular trajectory of the vortex orbit, while Gilbert damping acts to oppose this motion.

We are interested in the orbital motion of the vortex around the nanocontact, so it is more convenient to work with the equations of motion for the radial ($R$) and angular ($\varphi$) variables, which can be obtained from (\ref{eq:thiele}) by constructing the appropriate combinations of the coupled differential equations. We find
\begin{eqnarray}
\frac{\partial R}{\partial t} & = & -\frac{\kappa\alpha D}{G^2}|I|+ \frac{\sigma_2 I a^2}{R}, \\
\frac{\partial \varphi}{\partial t} & = & \frac{(\alpha-\beta)D}{G} \frac{\sigma_2 I a^2}{R^2} + \frac{\kappa}{G}\frac{|I|}{R}.
\label{eq:thielephase}
\end{eqnarray}
The stationary solution is obtained by setting the time variation in the radial variable to zero, i.e., $\partial_t R = 0$. From this condition, a solution to the stationary orbit radius $R(t) = R_0$ is found,
\begin{equation}
\begin{split}
R_0 &= \frac{G^2 \sigma_2 a^2}{\kappa \alpha D}\sgn(I), \\
&= P \frac{\hbar}{2 e} \left( \frac{1}{\alpha \mu_0 M_s d} \right) \frac{\left(p -\cos{\Theta_0}\right)^2}{(2+\sin^2{\Theta_0} \ln[L/b])\sin\Theta_0}.
\label{eq:radius}
\end{split}
\end{equation}
There are a number of important points worth noting here. First, a physical solution for the radius of the steady-state orbit, $R_0 > 0$, exists only for one current polarity, $I > 0$, which corresponds to the case where electrons flow \emph{outward} from the nanocontact. Intuitively, we can understand this as an outward ``pressure'' exerted by the spin torques that counterbalances the Gilbert damping and therefore prevents the vortex spiraling into the contact center, which corresponds to the position of energy minimum in the Zeeman energy. Second, the orbital radius is \emph{independent} of the applied current and the nanocontact size, and depends only on material parameters. As such, it also follows that there is \emph{no} critical current for self-sustained oscillations, in stark contrast to nanopillar spin-torque oscillators that involve large amplitude spin waves. The present theory therefore predicts the existence of vortex oscillations for any value of the applied electron current $I>0$, provided a vortex is \emph{already} present in the nanocontact system. The onset of oscillations in this scenario is therefore be determined by a threshold for vortex nucleation, rather than a supercritical Hopf bifurcation as in the case for nanopillar oscillators. Indeed, the experimental observation that vortex oscillations persist \emph{below} an onset current, after having been nucleated above this current,~\cite{Pufall:PRB:2007,Devolder:SPIE:2009} supports this idea.

From the stationary condition for the radial dynamics, the frequency of the vortex oscillation can be evaluated directly by substituting $R_0$ for $R$ in (\ref{eq:thielephase}). If we assume that the nonadiabatic torque is similar in magnitude to the adiabatic torque, $\beta \simeq \alpha$, which is found to be a good approximation for transitional metal ferromagnets,~\cite{Hayashi:PRL:2006,Burrowes:NP:2010} the first term on the right hand side of (\ref{eq:thielephase}) can be neglected and the expression for the frequency reduces to the simpler form
\begin{equation}
\omega \equiv \partial_t\varphi \simeq \frac{\kappa |I|}{G R_0}=  \frac{\mu_0 \gamma}{\pi} \left(\frac{\sin\Theta_0}{p- \cos\Theta_0} \right) \frac{|I|}{R_0},
\label{eq:freq}
\end{equation}
where $|I|/R_0$ plays the role of an effective magnetic field, with the sense of rotation is given by the vortex core polarization $p$. This is also an appealing result because it allows a quantitative estimate of the vortex orbital radius to be obtained readily from experiment: By measuring the slope of the frequency versus current curve in zero applied field $(\Theta_0 = \pi/2)$, a measure of $R_0$ can be obtained directly.

\section{Power spectrum}

\begin{figure}
\includegraphics[width=6cm]{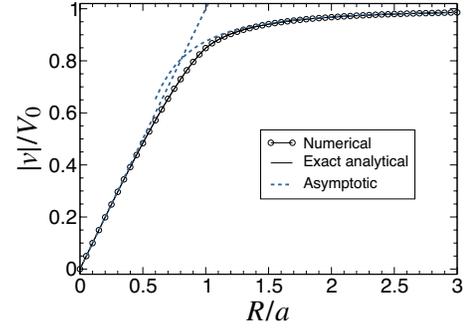}
\caption{\label{fig:GMRsignal_v_R}(Color online) Variation of the GMR voltage signal as a function of vortex radial distance, in normalized units. The variation, computed by numerical integration, is compared with the analytical result and simple asymptotic forms for small ($|v|/V_0 \simeq R/a$) and large amplitude (Eq.~\ref{eq:vgmr_largeorbit}) orbits.}
\end{figure}

The power spectrum measured in experiment is related to voltage variations $v(t)$ associated with the time-varying change in magnetoresistance in the nanocontact region. There are two contributions to the magnetoresistance: one from CPP currents, the other from CIP currents. Because the vortex orbital motion leads to the largest time-varying variation in magnetization in the contact region, the larger contribution to the total magnetoresistance is expected to come from the CPP component. Let $\Delta V$ represent the total CPP magnetoresistance for the spin valve. The time-varying component of interest can be written as
\begin{equation}
v(t) = \frac{1}{2} \Delta V \int_{\rm NC} d^3x \; \mathbf{m}(t) \cdot \mathbf{p}. 
\end{equation}
It is assumed that the fixed layer magnetization remains static and is uncoupled to the vortex dynamics in the free layer to simplify the calculations, although in practice this may need to be accounted for. To simplify the calculations, it will also be assumed that the fixed layer magnetization is uniform in the nanocontact region. As such, the voltage calculation amounts to averaging over the in-plane component of magnetization of the free layer over the point-contact region. Without loss of generality, it suffices to consider the $x$ component of magnetization in the free layer,
\begin{equation}
v(t) = \frac{1}{2} \Delta V d \int_{0}^{2\pi} d\phi \int_{0}^{a} dr \; r \frac{r \sin\phi - R \sin\varphi}{\sqrt{r^2 - 2 r R \cos(\phi-\varphi)+R^2}}.
\end{equation}
This integral can be solved using a power-series expansion in the variable $r/R < 1$ and summing over all terms. We find
\begin{equation}
v(t) = -\frac{1}{4} \pi a^2 d\; (\Delta V) \sin\varphi \;\; _2 F_1\left[ -\frac{1}{2};\frac{1}{2};2; (a/R)^2 \right],
\end{equation}
where $_2 F_1$ is a hypergeometric function which can expressed by the integral
\begin{equation}
_2 F_1(a;b;c;z) = \frac{\Gamma(b)}{\Gamma(c) \Gamma(c-b)} \int_{0}^{1} dt \; \frac{t^{b-1}(1-t)^{c-b-1}}{(1-tz)^a}.
\end{equation}
A plot of the variation of the magnetoresistance signal magnitude as a function of orbit radius is shown in Fig.~\ref{fig:GMRsignal_v_R}.
%
For small amplitude motion, the GMR signal exhibits a linear variation as a function in the radial distance. In the limit of large amplitude motion, the hypergeometric function can be approximated by $_2 F_1\left[ -\frac{1}{2};\frac{1}{2};2; (a/R)^2 \right] \simeq 1 - a^2/(8 R^2)$, which leads to a simple expression of the time-varying magnetoresistance signal
\begin{equation}
v(t) \simeq V_0 \left(1 - \frac{1}{2}\left[\frac{a}{2 R(t)}\right]^2 \right) e^{i \varphi(t)},
\label{eq:vgmr_largeorbit}
\end{equation}
where $V_0 \equiv \pi a^2 d \Delta V /2$. This result indicates that the vortex nanocontact oscillator can be considered as a phase oscillator to a very good approximation, with radial fluctuations that decrease like $1/R^2$. As such the power spectrum of the voltage fluctuations $S(\omega)$, 
\begin{equation}
S(\omega) = \int_{-\infty}^{\infty} dt \; K(t) e^{-i \omega t},
\end{equation}
which is the Fourier transform of the voltage autocorrelation function $K(t)= \langle v(t)v^{*}(0) \rangle$, is well described by the phase variance of the oscillator, $\Delta\varphi^2$,
\begin{equation}
K(t) = V_0^2 \exp\left[i \langle \varphi(t) \rangle \right] \exp\left[-\frac{1}{2}\Delta\varphi^2(t)\right],
\end{equation}
where $\Delta\varphi^2(t) = \langle \varphi(t)^2 \rangle - \langle \varphi(t) \rangle^2$.  Once the phase variance is known, the lineshape and linewidth of the power spectrum can be obtained directly.

The power spectrum of any oscillator is broadened at finite temperatures by thermal noise. We can describe the influence of such stochastic processes on the oscillator dynamics by including additional noise terms in the equations of motion (\ref{eq:thiele}), which we can express symbolically as the set of coupled Langevin equations~\cite{Kamppeter:PRB:1999,Kamppeter:EPJB:1999}
\begin{equation}
\frac{d \mathbf{X}}{d t} = \mathbf{v}(\mathbf{X}) + \sqrt{q}\; \mathbf{\eta}(t).
\label{eq:langevin}
\end{equation}
Note that the noise contribution enters as an additive term, rather than a multiplicative process (such as a random field in the Landau-Lifshitz equation); it has been shown elsewhere that such an additive noise is adequate for of vortex dynamics.~\cite{Kamppeter:PRB:1999,Kamppeter:EPJB:1999} $\mathbf{\eta} = (\eta_X,\eta_Y)$ is a two-component vector that represents a Gaussian white-noise forcing, which possesses the spectral properties
\begin{equation}
\langle \eta_{i}(t) \rangle = 0; \;\;\; \langle \eta_{i}(t) \eta_{j}(t') \rangle = 2\delta_{ij} \delta(t-t'),
\end{equation}
with $q$ representing the thermal noise amplitude. The choice of 
\begin{equation}
q = \frac{\alpha k_B T \gamma}{M_s d}
\label{eq:noise}
\end{equation}
ensures that the fluctuation-dissipation theorem is satisfied in the absence of spin-transfer torques.

Since we are interested in describing the fluctuations about the steady-state orbital motion, we linearize the Langevin equations about the stationary orbit $R_0$, $R(t)=R_0 +r(t)$, which allows us to obtain from (\ref{eq:langevin}), 
\begin{equation}
\frac{\partial r}{\partial t} = - \frac{\sigma_2 I a^2}{R_0^2} r + q \left(1-\frac{r}{R_0}\right) + \sqrt{q} \, \eta_R(t),
\end{equation}
where $\eta_{R,\varphi}$ possess the same spectral properties as $\eta_{X,Y}$. By analogy with spin torque oscillator theory, we identify a restoration rate $\Gamma_r = \sigma_2 I a^2/R_0^2$, which describes the rate at which fluctuations in the orbital radius are damped out. The radial fluctuations are independent of the phase fluctuations and are driven by white thermal noise, subject to a spurious drift. By using material parameters relevant for typical experiments, it is straightforward to show that the term in $q$ is small compared with the additive noise term proportional to $\sqrt{q}$, which allows the Langevin equation in $r(t)$ to be reduced to the simpler Ornstein-Uhlenbeck process,
\begin{equation}
\frac{\partial r}{\partial t} = - \Gamma_r r + \sqrt{q} \, \eta_R(t).
\end{equation}
The formal solution to this equation, $r(t) = c \exp(-\Gamma_r t) + \sqrt{q} \exp(-\Gamma_r t) \int^t d\tau \, \eta_R(\tau) \exp(\Gamma_r \tau) $, where $c$ is a constant, allows the two-time autocorrelation function for the radial fluctuation to be computed directly. In the limit of large times (at which the initial correlations are forgotten), we recover the known result 
\begin{equation}
\langle r(t) r(t') \rangle = \frac{q}{\Gamma_r}\exp\left(\Gamma_r |t-t'|\right).
\end{equation}

The linearized Langevin equation for the phase dynamics is
\begin{equation}
\frac{\partial \varphi}{\partial t} = \frac{\kappa |I|}{G R_0} \left(1-\frac{r(t)}{R_0}\right) + \sqrt{q} \, \eta_\varphi(t),
\end{equation}
where we have neglected a cross term in $r(t) \eta_\varphi(t)$. In contrast to the radial dynamics, which is uncoupled from the phase fluctuations, the phase variable is coupled to the radial fluctuations. As a result, the phase dynamics is driven by the radial fluctuations $r(t)$, which appears as colored noise, in additional to the additive white noise proportional $\sqrt{q} \eta_\varphi(t)$. From the formal solution to this differential equation, and by using the spectral noise properties for $\eta$, we obtain the phase variance to be
\begin{equation}
\frac{1}{2}\Delta\varphi(t)^2 = \frac{q}{R_0^2} \left( \left[ 1 + \left(\frac{\kappa |I|}{G R_0} \frac{1}{\Gamma_r}\right)^2 \right] |t| - \left(\frac{\kappa |I|}{G R_0} \right)^2 \frac{1-e^{-\Gamma_r t}}{\Gamma_r^3} \right).
\end{equation}
We can identify a nonlinearity parameter $\nu$,
\begin{equation}
\nu \equiv  \frac{\kappa |I|}{G R_0}\frac{1}{\Gamma_r} = \frac{G}{\alpha D}, 
\end{equation}
along with a ``linear'' linewidth parameter $\Delta \omega_0 = q/R_0^2$, which allows the expression for the phase variance to be simplified to the form,
\begin{equation}
\frac{1}{2}\Delta\varphi(t)^2 = \Delta \omega_0 \left[ \left( 1+\nu^2 \right) |t| - \frac{\nu^2}{\Gamma_r} \left(1-e^{-\Gamma_r t} \right)  \right].
\end{equation}
This expression is identical to the phase variance for a spin torque oscillator, which describes an inhomogeneous broadening of the spectral line due to radial (amplitude) fluctuations.~\cite{Tiberkevich:PRB:2008}

Two limiting cases appear as a result of this inhomogeneous broadening. In the ``low'' temperature limit in which the coherence time, $\tau_c$, of oscillations is much longer than the inverse of the restoration rate, $\Gamma_r \tau_c \gg 1$, the exponential term can be neglected and the phase variance is proportional to $|t|$. As such, the power spectrum in this limit is described by a Lorentzian lineshape with a full width at half maximum (FWHM) of
\begin{equation}
\Delta \omega_{\rm LT} = \Delta \omega_0 \left(1+\nu^2\right) = \frac{q}{R_0^2} \left(1+\nu^2\right).
\label{eq:lt_linewidth}
\end{equation}
Through the linear dependence of $\Delta \omega_0$ on the noise parameter $q$, the linewidth in this limit varies \emph{linearly} as a function of temperature and is inversely proportional to the square of the steady-state radius $R_0$. Because the orbital radius is independent of the applied current, the linewidth is also independent of the current in the low temperature limit, which is in stark contrast to conventional spin torque oscillators for which the supercriticality, which describes the ratio between the applied and threshold currents, is a primordial factor. For the present case the linewidth is determined primarily by material parameters. In the opposite ``high'' temperature limit in which the coherence time is much shorter than the inverse of the restoration rate, $\Gamma_r \tau_c \ll 1$, the exponential function in the phase variance can be expanded in a power series to give,
\begin{equation}
\frac{1}{2}\Delta \varphi(t)^2 \simeq \Delta \omega_0 \left( |t| + \frac{1}{2}\Gamma_r \nu^2 t^2  \right).
\end{equation}
If the nonlinearity $\nu$ is sufficiently large, the linear term in $|t|$ can be neglected and the power spectrum is described by a Gaussian lineshape with a FWHM of
\begin{equation}
\Delta\omega_{\rm HT} = 2 \sqrt{2 \ln 2} \, |\nu| \sqrt{\Delta \omega_0 \, \Gamma_r} \approx 2.35 \frac{|\nu|a\sqrt{q \, \sigma_2 I}}{R_0^2} .
\end{equation}
In this limit, the temperature dependence of the linewidth is of the form $T^{1/2}$, which is consistent with inhomogeneous broadening due to phase-amplitude coupling.~\cite{Tiberkevich:PRB:2008} Furthermore, the linewidth also acquires a square-root dependence on the applied current through the restoration rate, which is markedly different from the low temperature case. This gives a direct experimental means of identifying the temperature regime by measuring the current dependence of the linewidth.

\section{Discussion and comparison to experimental data}

The fundamental premise of this theory is that sub-GHz voltage oscillations observed in experimental nanocontact systems are due to the orbital motion of a \emph{single} magnetic vortex. The quasi-linear current dependence of the oscillation frequency provides the strongest evidence to support this hypothesis to date. However, there remain open questions concerning how the single vortex state is attained from an initially uniform magnetic ground state. This crucial issue is a problem of topology and can be discussed in terms of the Skyrmion number $Q$.~\footnote{The Skyrmion number is a topological charge and is defined by $Q = -(1/4\pi) \int d^2x \; \partial (\cos{\Theta,\Phi})/\partial (x,y)$. See Ref.~\onlinecite{Tretiakov:PRB:2007} for a detailed discussion.} A vortex or an antivortex possesses a half-integer Skyrmion charge of $Q = n p / 2$, where $n$ is the winding number, with $n=1$ for vortices and $n=-1$ for antivortices, and $p$ is the core polarization. As such, a system with a single vortex possesses a charge of $Q = 1/2$, while the uniform state has $Q = 0$.~\cite{Tretiakov:PRB:2007} In this light, the single vortex and uniform magnetic states are in different topological sectors, which means some physical process that does not conserve topological charge needs to occur during the nucleation. It is interesting to note that a vortex-antivortex pair with parallel core polarizations is in the same topological sector as the uniform state, while a vortex-antivortex pair with opposite core polarizations possesses $Q = \pm 1$. The latter case is interesting because it admits a rotating solution;~\cite{Komineas:2007,Komineas:PRL:2007} in the absence of damping, such a vortex-antivortex pair (or ``dipole'') rotates about its center-of-mass with a frequency of $\omega_d = 4 \gamma A / (M_s l^2)$, where $l$ is the separation between the vortices and $A$ is the exchange stiffness. Indeed, the possible existence of such rotating vortex dipoles in a nanocontact geometry were seen in a recent numerical study by Berkov and Gorn.~\cite{Berkov:PRB:2009} While such a solution is theoretically appealing because a uniform magnetic state is conserved far from the nanocontact, it is not immediately apparent how dynamics involving vortex dynamics can account for experimental observations. First, there is no obvious mechanism by which a linear current dependence of the oscillation frequency can be obtained. As we have shown in a recent study,~\cite{Devolder:APL:2010} the Oersted-Amp{\`e}re field gives rise to a Zeeman potential that drives the vortex pair apart, which would lead to a \emph{decrease} in the oscillation frequency with current. Second, a rotating dipole pair possesses a Skyrmion charge of $Q=\pm 1$, which is also topologically distinct from the uniform ground state. Therefore, as for the single vortex case, some charge non-conserving process needs to take place.

As discussed briefly in Section~\ref{sec:large_orbit}, we assert that in-plane spin-torques, rather than perpendicular-to-plane torques, should play a dominant role in the vortex dynamics in magnetic nanocontact systems under zero or low-applied fields. This hypothesis is strongly supported by a number of recent experiments in which vortex oscillations have been reproducibly initiated in the absence of any applied magnetic field.~\cite{Devolder:APL:2009,Manfrini:APL:2009,Devolder:APL:2010} Under such conditions the magnetization orientation of the reference magnetic layer should lie entirely in the film plane, so there is no reason to expect a large contribution from the $F_{\rm{CPP},\perp}$ term, which is necessary for the existence of self-sustained oscillations.~\cite{Mistral:PRL:2008} This point is well-illustrated by a recent experiment on vortex oscillations in magnetic nanopillars in which large out-of-plane fields are required.~\cite{Dussaux:NC:2010} While the dominant CIP adiabatic torques lead to the same functional form as the $F_{\rm{CPP},\perp}$ term in the reduced equations of motion, there is a crucial difference between the two: CIP torques act independently of the vortex core polarization, while the existence of vortex oscillations under CPP torques depend on relative orientation of the core polarization with respect to the perpendicular spin torque component.~\cite{Mistral:PRL:2008,Dussaux:NC:2010} Therefore, vortex core reversal would not restrict self-sustained oscillations under CIP torques but would lead to a damped oscillatory regime under purely CPP torques. In this light, it would be interesting to see whether such core reversal processes could be detectable in a time-resolved experiment.

The hypothesis of self-sustained oscillations driven by CIP torques might explain the ubiquitous presence of higher harmonics in the power spectrum. In this picture, the magnitude of the spin-torques due to lateral currents in the film plane determines the shape of the vortex orbit. A uniform radial flow has been assumed in the present work as a matter of simplicity, but the actual current distribution in an experimental system would certainly be more complex. As such, one could expect elliptical orbits to result, which would lead to significant contributions to the harmonic content of the power spectrum.

An important test of the model described here is the slope of the frequency versus current relation, 
\begin{equation}
\frac{\partial \omega}{\partial |I|} = \frac{\kappa}{G R_0} = \frac{\kappa^2 \alpha D}{G^3 \sigma_2 a^2},
\end{equation}
which, as discussed previously, is a material-dependent parameter. It is a quantity that can readily be extracted from experimental measurements and represents a robust characterization of the oscillator properties because the oscillation frequency, as opposed to other spectral parameters such as the linewidth or power, is a stable physical parameter that depends solely on the vortex dynamics and is independent of thermal fluctuations. By using the experimentally determined values of $\mu_0 M_s = 1.56$ T, $d = 3.5$ nm, $\alpha = 0.013$, $L = 10$ $\mu$m from Ref.~\onlinecite{Devolder:APL:2009}, and by assuming a spin polarization of $P = 0.5$ and $b = 10$ nm, we find a theoretical value of $\partial \omega/\partial |I| = 4.4$ MHz/mA, which is within a factor of two of the observed slope of 7.4 MHz/mA.~\cite{Devolder:APL:2009,Manfrini:APL:2009} For the low-field vortex oscillation studies of Refs.~\onlinecite{Pufall:PRB:2007} and \onlinecite{Keller:APL:2009}, a frequency slope of $\approx 30$ MHz/mA is predicted (by assuming $\mu_0 M_s = 0.8$ T, $d = 5$ nm, $\alpha = 0.01$, $L = 10$, $P = 0.5$, and $b = 10$ nm). This is a factor of three larger than the observed slope of $\approx$ 10 MHz/mA. While some uncertainty exists in the spin polarization $P$ of the applied currents and the vortex core radius $b$ in these experiments, the agreement between theory and experiment nevertheless remains relative good on a quantitative level for a simple model with no adjustable parameters.

Another important point of comparison is the spectral linewidth. Indeed, understanding the underlying physics governing linewidths is crucial for any potential oscillator applications. From the experiment described in Ref.~\onlinecite{Devolder:APL:2009}, it was shown that the autocorrelation function for the measured high-frequency voltage signals can be well-described by a decaying exponential function, $\langle v(t) v(0) \rangle \propto \exp{(-|t|/\tau_c)}$, where $\tau_c$ is the characteristic coherence time of the voltage oscillations. As discussed in the previous section, this form for the autocorrelation function indicates that the spectral lineshape is Lorentzian, with a FWHM given by $\Delta \omega = 2/\tau_c$. For an applied current of 18.7 mA in zero applied field ($\Theta_0 = \pi/2$), a coherence time of 140 ns was observed.~\cite{Devolder:APL:2009} From the results obtained for the low-temperature limit in the previous section, the present theory predicts the current-independent value of $\tau_c \approx 70$ ns with the same experimental parameters, which is in reasonable agreement with the experimental value. In a similar manner, a current-independent linewidth of $\Delta f \approx 2$ MHz is predicted for the experiment of Ref.~\onlinecite{Keller:APL:2009}, which is within a factor of three of the experimentally measured FWHM of 0.78 MHz for the oscillation mode at 128 MHz. The time-domain analysis performed for this particular experiment shows that the dominant contribution to the spectral linewidth is due to phase noise, so the present theory is applicable.

The discrepancies between the theoretical and experimental values of the frequency and linewidth might originate from the way the Oersted-Amp{\`e}re field and spin torques are computed in this theory. For instance, the spatial profile of the Oersted-Amp{\`e}re field has been computed by approximating the current flow through the nanocontact with the flow through an infinite cylindrical wire. On the other hand, it has also been assumed that a large component of this current flows laterally from the nanocontact because of how the contact pads are located. These assumptions are not compatible with one another, and as a consequence both the spin-torques and the Oersted-Amp{\`e}re field are overestimated in the theory. A simple way of correcting these estimates is to  include ad-hoc correction factors $\kappa_0 \leq 1$ and $\sigma_0 \leq 1$ into the definitions of the Zeeman energy and spin-torque parameters, respectively,
\begin{equation}
\kappa' = \kappa_0 \kappa, \;\;\; \sigma_2' = \sigma_0 \sigma_2.
\end{equation}
Following this line of reasoning, the frequency versus current slope and Lorentzian linewidth also acquire these correction factors,
\begin{equation}
\left(\frac{\partial \omega}{\partial |I|}\right)' = \frac{\kappa_0^2}{\sigma_0} \left(\frac{\partial \omega}{\partial |I|}\right), \;\;\; \Delta \omega_{\rm L}' = \left( \frac{\kappa_0}{\sigma_0} \right)^2 \Delta \omega_{\rm L},
\end{equation}
but with different functional forms. By setting the ratios between the theoretical and experimental values of the frequency and linewidth to unity, it is found empirically that $\kappa_0 \approx 0.4$ and $\sigma_0 \approx 0.3$. The empirical value of $\kappa_0$ suggests that a half-infinite cylindrical wire, for which one would expect a correction factor of 0.5, gives a better approximation for computing the Oersted-Amp{\`e}re field. It would be interesting to see whether more detailed finite-element calculations of the current flow in realistic geometries would give such correction factors.

\section{Summary}

In summary, a theory of the power spectrum of current-driven vortex oscillations in magnetic nanocontacts has been presented. The theory is based on a rigid-vortex model and the equations of motion describing the vortex dynamics have been derived and solved in the steady-state limit. In contrast to conventional spin torque nano-oscillators that involve large-angle magnetization precession, the self-oscillatory state in the nanocontact system is found to exist with the only condition being the existence of a vortex in the system. As such, the onset of oscillations does not involve a Hopf bifurcation and therefore no critical current is predicted. It is found that spin-torques due to current flow in the plane of the free magnetic layer are crucial for the existence of self-sustained oscillations. The oscillation frequency is found to vary linearly as a function of current, in accordance with experimental observations, with the function form
\begin{equation}
\omega = \frac{\kappa |I|}{G R_0}, \tag{\ref{eq:freq}}
\end{equation}
where the orbital radius $R_0$ is current-independent and depends only on material parameters,
\begin{equation}
R_0 = \frac{G^2 \sigma_2 a^2}{\kappa \alpha D} \sgn(I). \tag{\ref{eq:radius}}
\end{equation}
In the low-temperature limit in which the coherence time of the oscillations is greater than the restoration rate of the radial fluctuations, it is found that the power-spectrum is described by a Lorentzian lineshape with a linewidth (FWHM) of
\begin{equation}
\Delta\omega_{\rm LT } = \Delta\omega_0 \; (1+\nu^2), \tag{\ref{eq:lt_linewidth}}
\end{equation}
where $\Delta\omega_0$ is a ``linear'' linewidth that is proportional to the temperature and $\nu$ is a ``nonlinearity'' parameter that represents the ratio between the magnitude of the damping dyadic and the gyrovector, $\nu = \alpha D/G$. At higher temperatures, the lineshape is inhomogeneously broadened by fluctuations in the orbital radius and a Gaussian lineshape is predicted, with a current-dependent linewidth that varies like $\sqrt{I}$.

\begin{acknowledgments}
The authors thank R. E. Camley, M. Pufall, M. Keller, G. Hrkac, and C. Chappert for stimulating discussions. This work was partly supported by the European Communities program ``Structuring the ERA'', under Contract No. MRTN-CT-2006-035327 SPINSWITCH, the French National Research Agency (ANR), under contract no. VOICE PNANO-09-P231-36, and by the local government of R{\'e}gion Ile-de-France within the ``C'Nano IdF'' program.
\end{acknowledgments}

\bibliography{articles}

\end{document}